\documentclass{llncs} 

\usepackage{amsmath,amssymb,amsfonts,cases,graphicx,subfigure}
\usepackage{graphicx,url,bm,multirow,multicol}
\usepackage[usenames]{color}
\newcommand{\ignore}[1]{}
 
\newtheorem{insight}{\bf Insight}

\usepackage{url}
\usepackage[linesnumbered,ruled,vlined]{algorithm2e}
\SetKwInput{KwInput}{Input}                
\SetKwInput{KwOutput}{Output} 
\usepackage{diagbox}

\newcommand{\TL}{{\ensuremath{\sf TL}}}
\renewcommand{\L}{{\ensuremath{\sf L}}}

\begin{document}

\title{Smart Home Cyber Insurance Pricing}


\author{Xiaoyu Zhang\inst{1}
 \and 
Maochao Xu\inst{2}
\and 
Shouhuai Xu\inst{3}
}

\institute{Jiangsu Normal University, Xuzhou, Jiangsu, China\and
Illinois State University, Normal, IL, USA \and
University of Colorado Colorado Springs, Colorado Springs, CO, USA}

\maketitle

\begin{abstract}
Our homes are increasingly employing various kinds of Internet of Things (IoT) devices, leading to the notion of smart homes. While this trend brings convenience to our daily life, it also introduces cyber risks. To mitigate such risks, the demand for smart home {\em cyber insurance} has been growing rapidly. However, there are no studies on analyzing the {\em competency} of smart home cyber insurance policies offered by cyber insurance vendors (i.e., insurers), where `competency' means the insurer is profitable and smart home owners are not overly charged with premiums and/or deductibles. In this paper, we propose a novel framework for pricing smart home cyber insurance, which can be adopted by insurers in practice. Our case studies show, among other things, that insurers are over charging smart home owners in terms of premiums and deductibles.
\end{abstract}

\keywords{Smart home  cyber insurance \and cyber risk \and pricing strategy}

\section{Introduction}
The omnipresence of Internet of Things (IoT) technology allows average households to transform into smart homes to provide more convenient and comfortable living environments. According to a report by the { Zion Market Research \cite{Zion},} the global smart home market is likely to reach US\$137.9 billion by 2026.
This explains why smart homes are attracting increasing attention from 
cyber insurance vendors (i.e., insurers).  

\vspace{-2em}
\begin{figure}[!htbp]
\centering
\includegraphics[width=.5\textwidth]{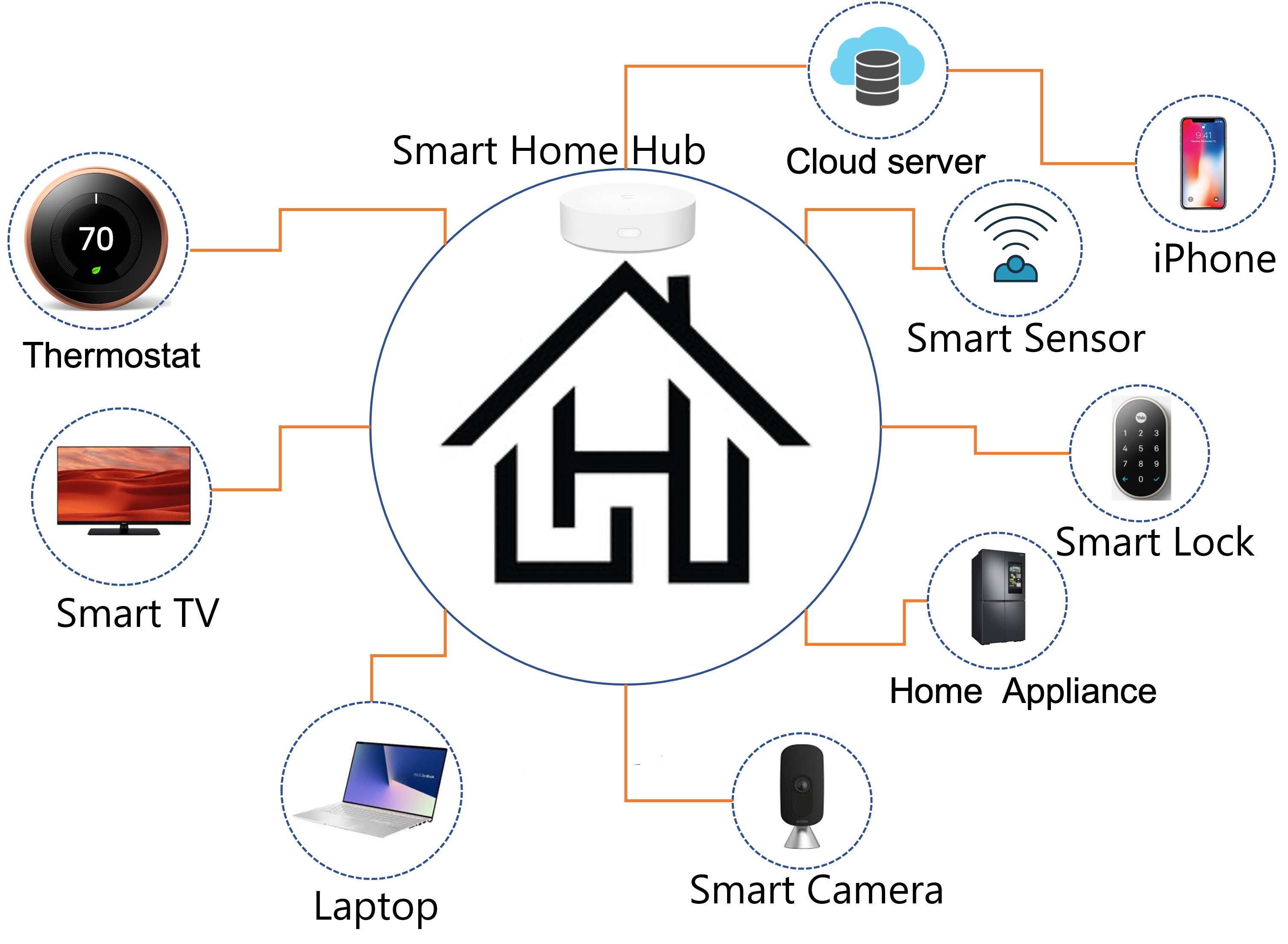}
\vspace{-1em}
\caption{Illustration of a smart home with 9 devices (excluding cloud server).\label{fig:sh}}
\end{figure}
\vspace{-2em}

Figure \ref{fig:sh} illustrates the smart devices in a smart home. As illustrated in Figure \ref{fig:sh}, there are a variety of IoT devices in a smart home 
\cite{marikyan2019systematic}, including thermostats, TVs, laptops, cameras, locks, sensors, smart home appliances, and alarms.
These devices may collect and exchange data with each other via a local 
network.
Moreover, there is a possibility of a home gateway that serves as a hub, 
and a cloud server may be involved in for storing bulk data for the long term.

While ushering in great convenience, these IoT devices, like many other new technologies, contain vulnerabilities that can be exploited by attackers to cause damages to smart home owners \cite{8835392,jacobsson2016risk}. For instance, an attacker can wage 
eavesdropping attacks 
to steal personal information, or exploit vulnerabilities in smart cameras to extract private videos for cyber extortion purposes \cite{bugeja2021prash}. 
As another example, the 2024 IOT Security Landscape Report \cite{bitdefender}, which is based on the threat intelligence sampled from 3.8 million smart homes around the world that are protected by the NETGEAR Armor powered by Bitdefender, shows that about 50 million IoT devices generate more than 9.1 billion security events related to vulnerabilities. These highlight that smart home networks have become a target of cyber attackers. 

To mitigate smart home cyber risks, insurance companies have expanded their services to include smart home cyber insurance. The current smart home insurance market typically provides coverage for smart home cyber insurance as an add-on to a standard home insurance policy. For instance, a {home safety} insurance \cite{valuepenguin} offers smart home cyber insurance coverage, including data breaches, computer and home systems attacks, cyber extortion, and online fraud, while coverage limit tops out at \$50,000 and a \$500 deductible; State Farm  \cite{valuepenguin} offers a personal cyber insurance add-on to its standard home insurance policy by covering smart home cyber risks, including cyber attacks and extortion, with a coverage limit of \$15,000; AIG \cite{valuepenguin} offers an insurance coverage up to \$250,000 at a \$1,652 annual premium. 



However, we are not aware of any study in the public literature the analyzes the {\em competency} of smart home cyber insurance, where competency indicates that smart home insurers are profitable and smart 
home owners do not overpay in premiums and/or deductibles. This is crucial to foster a viable and sustainable smart home cyber insurance market because home owners have no idea whether they are charged and covered appropriately, while noting that insurers would never make their proprietary pricing methods public. 
This highlights the importance of studying principled solutions to the problem of smart home cyber insurance pricing.
To our knowledge, there are essentially no studies on smart home cyber insurance despite the many studies on smart home cyber security
(e.g., \cite{kozlov2012security,lee2014securing,jacobsson2016risk,8835392,marikyan2019systematic}). In this paper, we make a first step to fill the void.

\smallskip

\noindent{\bf Our contributions}. 
In this paper, we make two contributions. First, we propose a novel framework for tackling the smart home cyber insurance problem. 
The framework aims to identify {\em competent} smart home cyber insurance policies where insurers are profitable and smart home owners do not overpay in premiums and/or deductibles. The framework considers smart home cyber risks in terms of {\em business lines}, which are often used by the insurance sector. This framework can be adopted by insurers to price smart home cyber insurance premiums and deductibles.  
Second, we demonstrate the usefulness of the framework by conducting case studies. Our findings include: (i) the current smart home cyber insurance requiring deductibles overly charges smart home owners; (ii) the current smart home cyber insurance requiring no deductibles is not profitable; and (iii) our framework leads to more competent smart home cyber insurance market, meaning that insurers remain profitable while home owners do not overpay in premiums or deductibles.


\smallskip

\noindent{\bf Related work}. 
To our knowledge, the present study is the first on smart home cyber insurance, which has unique aspects, such as prevalence of IoT devices and risks of cyber extortion (via private video), online fraud, and property theft.
This is true despite the many studies on smart home devices (e.g., \cite{jacobsson2016risk,denning2013computer,lee2014securing,8835392,das2011home,kozlov2012security,XuUsenixSecurity2023}) and cybersecurity risk management and dynamics (e.g., \cite{XuCybersecurityDynamicsHotSoS14,XuBookChapterCD2019,XuMTD2020,Pendleton16,XuSTRAM2018ACMCSUR,XuSciSec2021SARR}). 
Nevertheless, cyber insurance 
has been studied in other contexts (than smart homes) in two categories: {\em model-driven} vs. {\em data-driven}.
Model-driven studies include:
modeling the cyber insurance market
and pricing \cite{bohme2010modeling} via the {\em expectation principle} (one of the 4 principles we will consider);  
insurance pricing \cite{xu2019cybersecurity} via the {\em standard deviation} principle (another of the 4 principles we will consider);
we refer to  \cite{cremer2022cyber,awiszus2023modeling,he2024modeling} for excellent surveys.  
The present study falls into this category, but initiating the investigation of smart home cyber insurance. 
Moreover, the present study appears to be the first that investigates cyber risks based on {\em business lines}, making academic research one step closer to insurance practice. 
On the other hand, data-driven cyber insurance pricing has been investigated in \cite{sun2021modeling,eling2022unraveling,ma2022frequency,sun2023multivariate}.

\vspace{-2em}

\begin{table}[htbp!]
    \centering
    \begin{tabular}{|c|l|}
    \hline
    Notation & \multicolumn{1}{|c|}{Description} \\
    \hline
    \hline
    $n$ & Number of vulnerabilities \\
    \hline
    $M$ & Number of risk business lines\\
    \hline
    $V$ & Set of nodes(vertices) with each representing a vulnerability\\
    \hline
    ${\bf S}$ &  Set of random variables with each representing a vulnerability state \\
    \hline
    $\mathbb{S}$ & Set of all possible vulnerability states\\
    \hline
    $E$ & Set of directed edges where $(i,j)\in E$ indicating vulnerability $i$ can exploit $j$\\
    \hline
    $G(V,E)$ & Directed vulnerability graph consist of $V$ and $E$ \\
    \hline
    ${\bf pa}_j$ & Parent node set of vulnerability $j \in V$\\  
    \hline
    ${\rm L_m}$  & Loss in business lines $m \in \{1, \ldots,M\}$ without insurance\\
    \hline
    $\mathcal{L}_m$ & Subset of $V$ whose exploitation can incur loss ${\rm L_m}, m \in \{ 1,\ldots, M\}$ \\
    \hline
   $X_m$ & Loss in business line $m\in \{1,\ldots,M\}$ with insurance\\
    \hline
    ${\rm TL}$ & Total loss in $M$ business lines\\
   \hline
   $(d,C)$ & Deductible and coverage limit in an insurance product \\ 
   \hline
   $\theta$ &  Parameter reflecting
the risk attitude of the insurer \\
\hline
$p_i$ & Exploitation probability(EPSS score) of
entry point $i \in V$ \\
\hline
    $e_{ij}$ & Conditional probability that
vulnerability $i \in V$ exploited $j \in V$\\
    \hline
    \end{tabular}
    \caption{Summary of the major notations used in this paper.}
    \label{tab:notation}
\end{table}

\vspace{-2em}

\noindent{\bf Paper outline}. Section \ref{sec:framework} presents the framework. 
Section \ref{sec:case} conducts case studies on pricing.
Section \ref{sec:conclusion} concludes the paper with future research directions. {For the easy of reference, Table \ref{tab:notation}
summarizes the main notations used in this paper.}

\section{Framework for Smart Home Cyber Insurance Pricing}
\label{sec:framework}

Our framework consists of four steps: (i) defining smart home cyber insurance business lines; (ii) identifying and representing smart home cyber risks; (iii) {modeling smart home cyber risks in terms of business lines}; and (iv) determining smart home cyber insurance premiums and deductibles. 

\subsection{Defining Smart Home Cyber Insurance Business Lines}

For cyber insurance purposes, an  insurer needs to estimate the potential loss of a smart home incurred by cyber attacks
in terms of {\em business lines}, which is an insurance term describing products offered by an insurer. In practice, the following
six business lines are widely used and thus adopted in this study. 
\begin{itemize}
\item {\em Data breach}  ($\L_1$). This { business line of risk} refers to the exposure of private data in a
smart home, such as the home owner's daily activities, emotions, health conditions, audios, and videos.
The data breach insurance covers attorney cost, IT professionals, and mitigation of damage.
 
\item {\em Loss of use}  ($\L_2$). This { business line of risk} refers to the damage incurred by the unavailability of {service. The insurance covers the cost associated with the recovery of service, including ``cleaning up'' compromised devices, data recovery, home applicant repair, and system restoration.}


\item {\em Ransomware}  ($\L_3$). This {business line of risk} refers to the loss incurred by ransomware, which encrypts the data on victims' devices.
{The insurance covers the ransom upon the approval of the insurer when no other methods can recover the data. Note that this risk is different from the {\em loss of use} risk because the only solution in this case is to pay the ransom.} 

\item {\em Cyber extortion}  ($\L_4$).  This {business line of risk} refers to when the attacker threatens to release sensitive personal data, activities, conversations, or videos of a victim,  the insurance covers what is being  
demanded by the attacker.

\item {\em Online Fraud}  ($\L_5$). This {business line of risk} refers to the financial loss incurred by cyber attacks that stole funds via unauthorized use of bank or credit cards, phishing schemes, and other types of fraud. 
The insurance covers the direct financial loss incurred by the attack.
 
\item {\em Property theft}   ($\L_6$). This {business line of risk} refers to the loss incurred by cyber attacks {against the cyber defense
systems employed at a smart home. The insurance covers the failure of the employed cyber defense system. Note that a smart home that employed cyber defense tools but suffered from a data breach can make claims with respect to $\L_1$ and $\L_6$. }
\end{itemize}

\subsection{Identifying and Representing Smart Home Cyber Risks}
\label{sec:risk-ident}


In practice, vulnerabilities in smart home devices can be detected by vulnerability scanners, such as Nessus \cite{Nessus} and OpenVAS \cite{OpenVAS}. It would be ideal to patch all vulnerabilities in smart homes to mitigate risks.  However, this is not always possible in practice (e.g., home owners do not have the technical expertise). As a consequence, vulnerabilities are present in smart home devices and can be exploited by attackers. Thus, we must deal with the presence of vulnerabilities in smart homes.

Each vulnerability is identified by a Common Vulnerabilities and Exposures (CVE) number and described by the Common Vulnerability Scoring Systems (CVSS) \cite{CVSS}. 
For each vulnerability, the base CVSS scores, also known as impact metrics, describe the impact when a vulnerability is exploited in terms of confidentiality, integrity, and availability;
the Exploit Prediction Scoring System (EPSS) score 
\cite{jacobs2021exploit} measures the probability that the vulnerability will be exploited within period of time after its public disclosure, {independent of others}. Thus, CVSS and EPSS together can describe the risk associated with a vulnerability, {assuming their exploitations are independent of each other}. That is,
the EPSS score can be seen as the probability of exploiting a  vulnerability as an entry point for penetrating into a smart home.

However, vulnerabilities are not necessarily exploited independent of each other; rather, the exploitation of one vulnerability often leads to the exploitation of another. Thus, we need to estimate the exploitation probability that goes beyond what is provided by EPSS. This requires us to describe how vulnerabilities may be exploited by attackers after gaining entry point into a smart home. 
For this purpose, we propose using a Graph-Theoretic representation as illustrated by the following example. 

Suppose a smart home has three vulnerabilities, say 
CVE-2022-22667 (in iPhone),  CVE-2018-3919 (in smart home hub), and CVE-2021-32934 (in smart camera). These  vulnerabilities enable the attacker to conduct the following attack: the attacker first exploits CVE-2022-22667 in the iPhone,
then pivots from the compromised iPhone to exploit CVE-2018-3919 to attack
the smart home hub, then pivots from the compromised hub to exploit CVE-2021-32934 to attack the smart camera, and finally breaches videos taken by the camera.


The preceding intuitive discussion prompts us to propose the following steps for identifying and representing smart home cyber risks. 
\begin{enumerate}
\item Scan vulnerabilities in smart home devices. The detected vulnerabilities are represented by their CVE numbers, CVSS scores, and EPSS scores.  
\item Create a directed vulnerability graph
$G=(V,E)$, where $V$ is the set of nodes (i.e., vertices) {with each representing a vulnerability} and $E$ is the set of arcs (i.e., directed edges) with $(i,j)\in E$ representing that the exploitation of vulnerability $i\in V$ could lead to the exploitation of vulnerability $j\in V$. For instance, the aforementioned attack scenario can be represented as a vulnerability graph $G$, where $V=\{1,2,3\}$ (respectively representing CVE-2022-22667, CVE-2020-27403, and CVE-2018-3919) and $E=\{(1,2),(2,3)\}$.
\end{enumerate}




\ignore{

 \begin{figure}[!htbp]
\centering
\includegraphics[width=.9\textwidth]{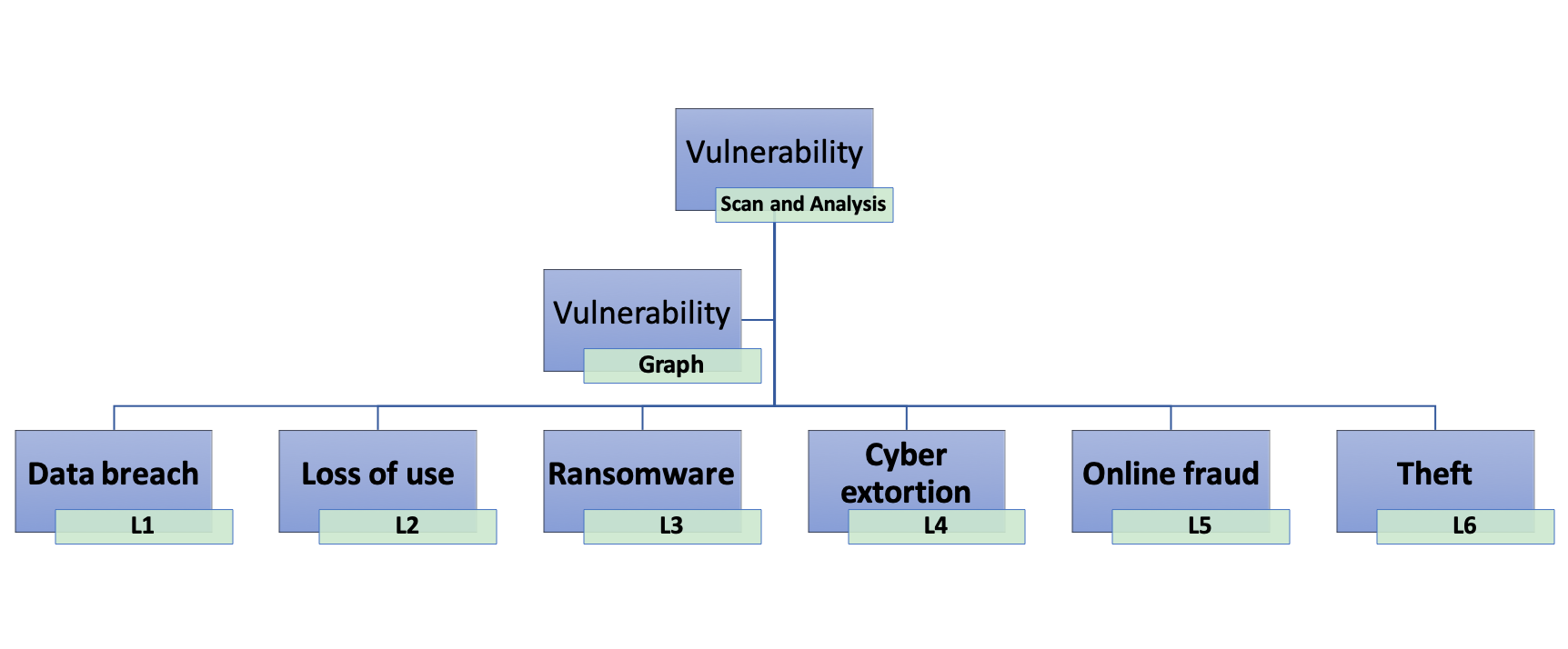}
\caption{Various insurance risks in a smart home ecosystem. \label{fig:fog-net}}
\end{figure}

Figure \ref{fig:fog-net} shows those risk categories from the perspectives of insurance companies based on the vulnerability graph.  

}

Smart home owners typically lack the expertise to create vulnerability graphs. Therefore, insurers should assess cyber risks by understanding attack paths and developing these vulnerability graphs. This can be achieved, for instance, by collaborating with a third-party smart home cybersecurity assessment service provider and/or utilizing an insurer's own cybersecurity team.

\subsection{Modeling Cyber Risks in Terms of Business Lines}

For cyber insurance purposes, we need to estimate the Total Loss (\TL), which is a random variable and incurred by cyber attacks against devices in a smart home. Suppose there are $M$ business lines of risk, represented by $\L_1,\ldots, \L_M$. In order to estimate $\TL$ or its distribution $\Pr\left(\TL\le \ell \right)$, we need to estimate the random variable of loss per business line, namely $\Pr\left(\L_m\le \ell_m \right)$ for $m\in \{1,\ldots,M\}$ where 
$\Pr\left(\TL\le \ell \right)=\Pr\left(\sum_{m=1}^M \L_m\le \ell \right)$.
To estimate $\L_1,\ldots, \L_M$ and thus $\TL$, we need to model the security states of smart home devices because the compromise of different devices may incur losses in different business lines. That is, we need to model the exploitation probability for {\em each and every} vulnerability $i\in V$ in the vulnerability graph $G=(V,E)$ produced in the previous step. 

For a vulnerability that is exploited as an entry point into a smart home (e.g., vulnerability in the iPhone in the preceding example), the EPSS score, or a justifiable amendment of it, can be used as its exploitation probability.
However, determining the exploitation probability of a non-entry vulnerability in the smart home would require a careful treatment. For instance, the exploitation probability of the smart home hub in the preceding example 
would conditionally 
depend on the exploitation of the iPhone, 
and the exploitation of the smart home camera would conditionally depend on the exploitation of the smart home hub. Given vulnerability graph $G=(V,E)$, there are methods to determine the exploitation probabilities of the non-entry nodes, such as the method that will be used in our case study. 

To make our framework able to accommodate any relevant method, we only require that a suitable method, which is equivalent to estimating the security state of each node in a smart home, should take the following as input: (i) A vulnerability graph $G=(V,E)$, where $V$ is the set of nodes with each representing a vulnerability and $E$ is the set of arcs with $(i,j)\in E$ meaning that the exploitation of $i\in V$ could lead to the exploitation of $j\in V$; (ii) each $i\in V$ is annotated with a CVE number, the CVSS scores of $i$, and the EPSS score of $i$; and (iii) each arc $(i,j)\in E$ is annotated with the condition probability that $j\in V$ is exploited when $i\in V$ is exploited.

To demonstrate feasibility, Algorithm \ref{alg:alg1} simulates the distributions of $\L_m$ and $\TL$ on inputs: (i) the set $V$ of smart home devices; (ii) the exploitation probability $p_i$ of device $i\in V$; (iii) the subset $\mathcal{L}_m \subseteq V$ of nodes whose exploitation can incur loss $\L_m$ in business line $m$, where $m\in \{1,\ldots,M\}$; (iv) the conditional probability $e_{ij}$ for $i,j\in \{1,\ldots,n\}$; and (v) the number $R$ of simulation runs.

\ignore{
. {\color{red}And distribution of the $\L_m$ when vulnerabilities in $\mathcal{L}_m$ are exploited, where $m\in \{1,\ldots,M\}$.}\footnote{{\color{blue}See in section 3.2 for example, we assumed that  $L_2 \sim \exp(\lambda_2)$, where $\lambda_2=1/640 V_3 + 1/320 V_5$, thus, if we know the state of $V_3, V_5$, we can explicitly obtain condition distribution: $P(L_2<l_2|V_3,V_5)$, then we can draw loss samples from it. (meaning in Algorithm 1 we should let ``distribution of the ${\rm L_m}$......" be input)}}
}

\begin{algorithm}[!htbp]      
\label{alg:alg1}
\caption{Simulating distributions of $\L_m$ and $\TL$}
\KwInput{$V=\{1,\ldots, n\}$; 
the exploitation probability $p_i$  (e.g., EPSS score) of each possible entry point $i\in V$; $\mathcal{L}_m$; distribution of the $\L_m$ when vulnerabilities in $\mathcal{L}_m$ are exploited, where $m\in \{1,\ldots,M\}$; conditional probability $e_{ij}$ for $ i, j\in \{1, \ldots, n\}$; the number $R$ of simulation runs}

\KwOutput{Simulated distributions of $\L_m$ for $m\in \{1,\ldots,M\}$ and total loss $\TL$}

{Draw 
$G=(V,E)$ where $V=\{1,\ldots, n\}$ and $(i,j)\in E$ means 
the exploitation of vulnerability $i\in V$ can lead to the exploitation of $j\in V$}


\For{$r= 1$ \KwTo $R$}{
{Generate Bernoulli vector ${\bm S}^{(r)}=(S_1,\ldots,S_n)$ based on the $p_{i}$'s of the possible entry points 
and $e_{ij}$ for $i,j\in \{1, \ldots, n\}$}

{Determine the $\L_m$'s that are impacted by the exploitation of $i\in V$}

\For{$m= 1$ \KwTo $M$}{
{Randomly generate loss $\L^{(r)}_{m}$ 
according to the distribution  of $\L_m$  when the vulnerabilities in $\mathcal{L}_m$ are exploited 
}
}

{Record $\L^{(r)}_m$ and $\TL^{(r)}=\sum_{m=1}^M \L^{(r)}_m$ for $m\in \{1,\ldots,M\}$ and $r\in \{1,\ldots, R\}$}
}

\Return{$\L^{(r)}_m$ for $m\in \{1,\ldots,M\}$ and total loss $\TL^{(r)}$ where $r\in \{1,\ldots, R\}$}
\end{algorithm}

\subsection{Determining Premiums and Deductibles}
\label{sec:premium}
Given the estimated loss in business line $\L_m$ for $m\in \{1,\ldots,M\}$, this step determines the cyber insurance premium {and deductible}.
Let $C$ be the coverage limit (i.e., the maximum pay by cyber insurer to a smart home owner in the case of successful cyber attacks against the smart home), which is often an input parameter determined by the insurer.
Let $d$ denote the deductible of a smart home owner, which is also an input parameter and determined by the insurer. Since an insurer does not necessarily know in advance about what a suitable deductible $d$ should be, the insurer would need to try a range of candidate deductible $d$'s based on the following two pre-determined parameters, which are widely used in the insurance industry \cite{klugman2012loss}: 
the Profit to the insurer, defined as $\mbox{Profit}=\mbox{Premium} - \mbox{Claim}$, {where Premium is the total amount of premiums collected by the insurer and Claim is the total amount of claims paid to the smart home owners by the insurer}; and the Loss Ratio (LR) to the insurer, defined as
$\mbox{LR}= \mbox{Claim}/\mbox{Premium}$.
For example, the permissible LR may be 40\% to assure that an insurer can make a profit.

To an insurer, the loss in business line $m$ is defined as
\begin{equation}\label{eq:remain}
 {\rm X}_m=\min\{({\L}_m-d)_+,C\}, 
 \end{equation}
where 
$(\L_m-d)_+=\L_m-d$ if $\L_m>d$ and $(\L_m-d)_+=0$ otherwise.
In Actuarial Science, there are many premium pricing principles \cite{kaas2008modern}. In the present study, we  consider the following 4 popular premium pricing principles. 

\smallskip

\noindent{\bf Expectation principle} \cite{kaas2008modern}: Under this principle, the premium is defined as $\rho_1({\rm X}_m)$ $=(1+\theta)\,E({\rm X}_m)$, 
where 
$E$ is the expectation function, and $\theta$ 
reflects the risk attitude
of the insurer (i.e., $\theta = 0$ means risk-neutral, $\theta < 0$ means risk-seeking, and $\theta > 0$ means risk-averse). The basic idea behind the principle is to adjust the expected value of the loss by a factor that accounts for the insurer's risk attitude. The principle is easy to understand and implement. However, it has the disadvantage that it does not account for the variability in the distribution of loss ${\rm X}_m$, which is critical when dealing with highly uncertain events (e.g., which vulnerabilities will be exploited).

\smallskip

\noindent{\bf Standard deviation principle} \cite{kaas2008modern}: Under this principle, the premium is defined as $ \rho_2({\rm X}_m)=E({\rm X}_m)+\theta\, \sqrt{{\rm Var}({\rm X}_m)}$,
where $E$ is the same as above (i.e., expectation), ${\rm Var}$ is the variance function, and $\theta$ 
reflects the risk attitude 
of the insurer (same as above). The basic idea behind the principle is to adjust the expected value of the loss by a factor proportional to the standard deviation of the loss, thus accounting for the variability in the distribution of loss ${\rm X}_m$. This leads to the advantage of accommodating uncertainty of the loss distribution, offering a more accurate pricing. Its disadvantage is that it assumes a linear relationship between {the loss ${\rm X}_m$ and 
and its standard deviation, which may not be true in some circumstance (e.g., when losses follow a highly skewed   distribution).}

\smallskip

\noindent{\bf Gini mean difference (GMD) principle} \cite{furman2017gini,furman2019computing}:  Under this principle, the premium is defined as  $\rho_3({\rm X}_m)=E({\rm X}_m)+\theta\, 
E (|{\rm X}_{m1}-{\rm X}_{m2}|)$,
where {$\theta$ also reflect an insurer's risk attitude (as above),}
$E (|{\rm X}_{m1}-{\rm X}_{m2}|)$
is the Gini Mean Difference (GMD) that measures the statistical variability between a pair of independent realizations
of ${\rm X}_m$, denoted by ${\rm X}_{m1}$ and ${\rm X}_{m2}$.
The basic idea behind the principle is to adjust the expected value of the loss by a factor proportional to the GMD, which captures the average absolute difference between pairs of independent realization of the loss, thereby accounting for variability.
The advantage is that it is sensitive to the shape of the loss distribution and captures more information about the variability of the loss than the standard deviation. The disadvantage is that it is somewhat involved to cope with.


\smallskip

\noindent{\bf Conditional tail expectation {principle}} \cite{hardy2006introduction,tasche2002expected}: Under this principle, the premium is defined as $\rho_4({\rm X}_m)= E( {\rm X}_m| {\rm X}_m \geq {\rm VaR}_\beta )$,
where  
${\rm VaR}_\beta$ is 
the value-at-risk at level $\beta \in (0,1)$, namely ${\rm VaR}_\beta=\min_\gamma\left\{ \gamma: \Pr\left( {\rm X}_m \leq \gamma \right) \geq \beta \right\}$. The basic idea behind the principle is to determine the premium based on the conditional expectation of the loss ${\rm X}_m$, given that the loss exceeds the threshold that is defined by the value-at-risk at a desired confidence level $\beta$. Its advantage is that it focuses on the tail of the loss distribution, which is important for assessing extreme risks and ensuring that sufficient funds are available to cover significant losses. Its disadvantage is that it only considers losses beyond a certain threshold and ignores the distribution of losses below this threshold, which may offer an incomplete assessment of the overall risk.


Given that each of the preceding principles has its advantages and disadvantages, the overall guideline in selecting principles to guide pricing is the following: choose a principle that aligns with the specific risk management objective, the nature of the risks being insured, and the risk attitude of an insurer.   For our case study, we compare all the principles to understand their implications and suitability for different scenarios {in smart home insurance for the first time.}  

\section{Case Studies}\label{sec:case}



We inherit the 6 business lines defined in the framework. Thus, in what follows, we only focus on the other three steps. 

\subsection{Identifying and Representing Smart Home Cyber Risks}

Our case study is based on the smart home illustrated in Figure \ref{fig:sh}, which has 9 smart home devices. 
Suppose a vulnerability scanning process shows that 7 (out of the 9) devices 
contain the vulnerabilities listed in Table \ref{Tab:V-Loss-sh}.
Thus, the vulnerability graph $G=(V,E)$ has $V=\{1,2,3,4,5,6,7\}$,
where node 1 supposedly represents vulnerability CVE-2022-22667 in the iPhone, node 2 for CVE-2020-27403 in the smart TV, node 3 for CVE-2018-3919 in the smart home hub, node 4 for CVE-2021-29438 in the smart sensor, node 5 for CVE-2021-32934 in the smart camera, node 6 for CVE-2019-7256 in the smart lock, and node 7 for CVE-2017-8759 in the laptop.

\vspace{-2em}
\begin{figure}[!htbp]
\centering
\subfigure[Attack scenarios]{\includegraphics[width=.44\textwidth]{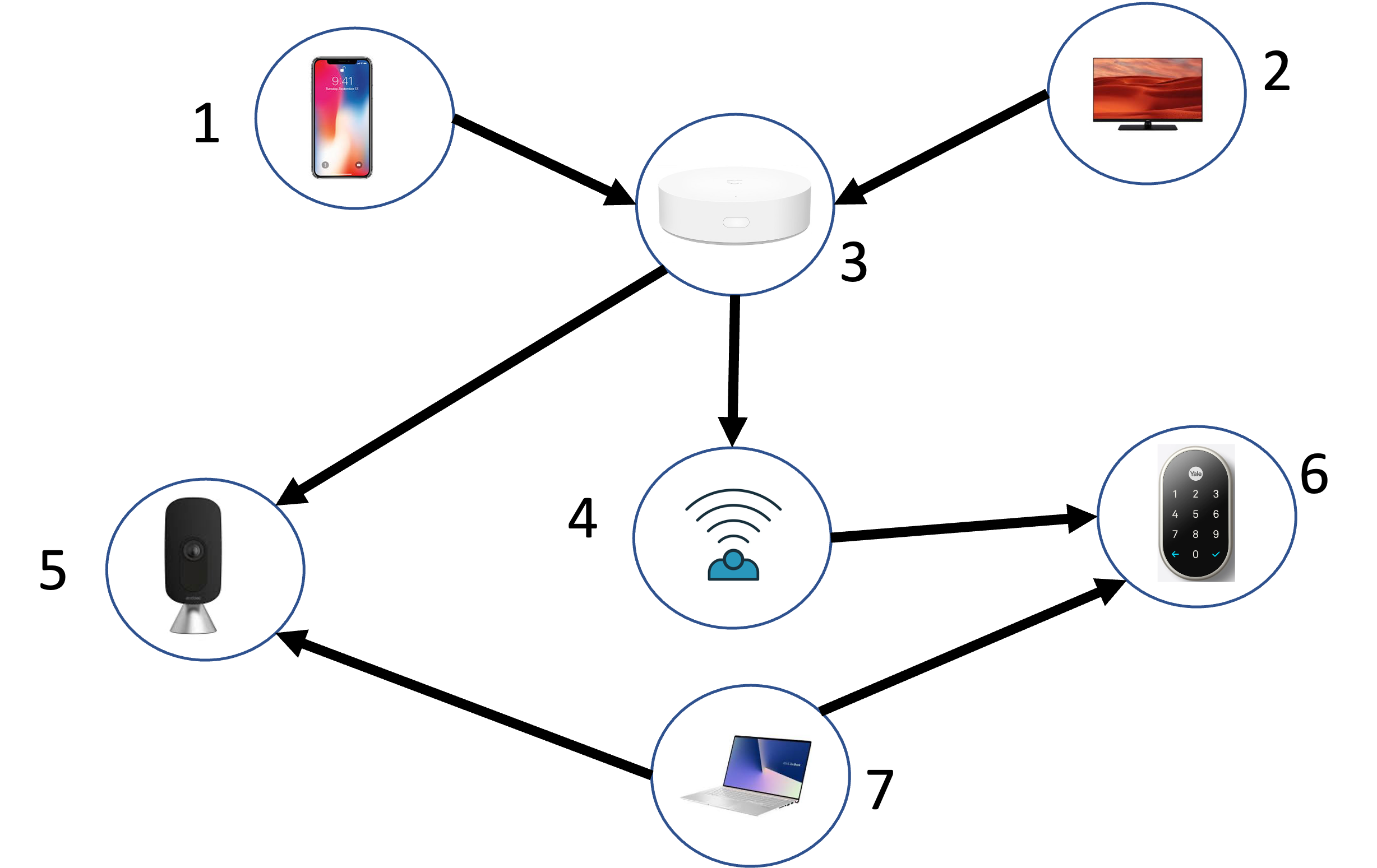}\label{fig:attack}}\quad
\subfigure[{Bayesian Attack Graph (BAG)}]{\includegraphics[width=.4\textwidth]{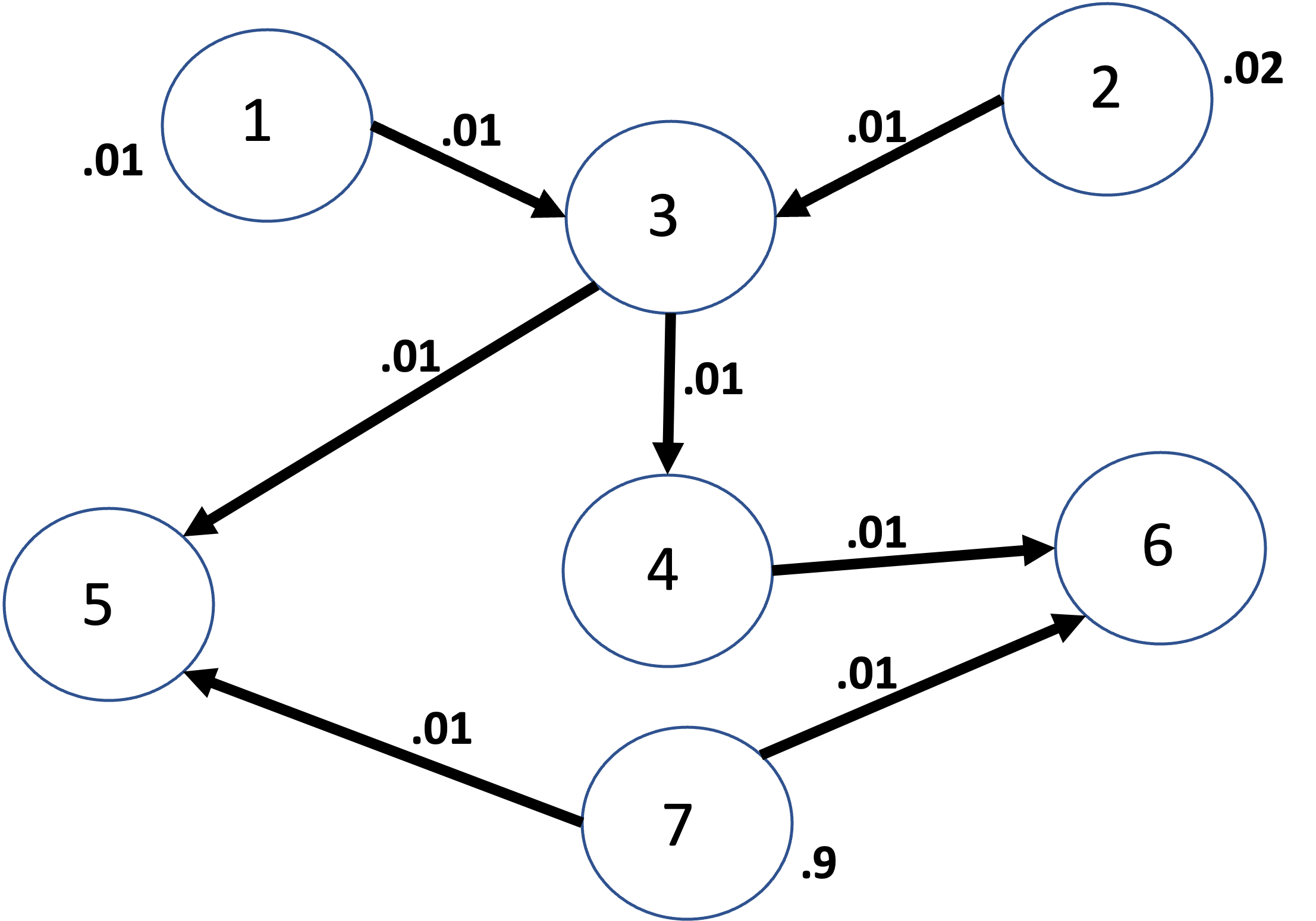}\label{fig:sh_new}}
\vspace{-1em}
\caption{{Graph-theoretic representation of vulnerabilities and attacks in a smart home. (a) Attack steps represented as arcs.
(b) BAG with exploitation probabilities, such as {$\Pr(S_7=1)=.9$ and $\Pr(S_5=1|S_7=1)=.01$}. 
\label{fig:graph-theoretic-representation}}
}
\end{figure}
\vspace{-2em}

Figure \ref{fig:attack} illustrates the Graph-Theoretic representation of vulnerabilities and attacks in the smart home based on the following assumptions.  (i) Nodes 1, 2, and 7 are the entry points, meaning that they (i.e., the devices containing these vulnerabilities) can be exploited by the attacker from outside of the smart home. (ii) Arc $(i,j)\in E$ means that compromise of node $i\in V$ can cause the compromise of node $j\in V$. Note that there are 6 attack paths, namely
$1\to 3 \to 5$; $1\to 3 \to 4 \to 6$; 
$2 \to 3\to 5$; $2\to 3 \to 4 \to 6$; $7 \to 5$; and $7\to 6$.
For example, attack path $1\to 3 \to 5$ means the following: 
the attacker first compromises the iPhone, then pivots to compromise the smart home hub, and finally pivots to compromise the smart home camera. 
This attack may cause, for example, {\em data breach} ($\L_1$), {\em online fraud} ($\L_5$),
and {\em cyber extortion} ($\L_4$) because the attacker targets data stored in the smart home hub or recorded at the smart home camera.
Similarly, attack path $1 \to 3 \to 4 \to 6$ can cause
risks including {\em data breach} ($\L_1$) and {\em property theft} ($\L_6$) because the attacker can unlock the door; attack path $7\to 5$
can cause risks including {\em loss of use} ($\L_2$) and {\em cyber extortion} ($\L_4$); and attack path $7\to 6$
can cause risks including {\em data breach} ($\L_1$), {\em ransomware} ($\L_3$) and {\em property theft} ($\L_6$);


In our case study we use the Bayesian Attack Graph (BAG) approach \cite{koller2009probabilistic,poolsappasit2011dynamic} to model cyber risks because of its simplicity.
At a high level, BAGs are graphical models representing information about network vulnerabilities and how they may be exploited in terms of attack paths. 
This leads to the BAG illustrated in Figure \ref{fig:sh_new}, where entry nodes 1 (iPhone), 2 (smart TV), and 7 (laptop) 
are assumed to be exploited by the attacker from outside of the smart home with the probability that is specified by their respective EPSS score, $.01$, $.02$, and $.9$. The probability $e_{ij}$ associated with arc $(i,j)\in E$ is defined as the conditional probability that the exploitation of $i\in V$ leads to the exploitation of $j\in V$.
For simplicity,
we assume $e_{ij}=.01$ for all $(i,j)\in E$.

To make the discussion below useful in general cases, we consider $V=\{1,\ldots,n\}$ and $|V|=n$, with $n=7$ in the preceding example. 
Let $s_i=1$ denote the fact (i.e., with probability 1 or certainty) that vulnerability $i\in V$ is exploited (i.e., the corresponding device is compromised) and $s_i=0$ otherwise. Let $S_{i}$ be the random variable denoting $i\in V$ is exploited, where $S_i  =1$ means $i$ is exploited and $S_i  =0$ otherwise. With these notations, we can write $e_{ij}=\Pr(S_{j}=1|S_{i}=1)$. Let ${\bf S}=(S_1,\ldots,S_n)$, ${\bf s}=(s_1,\ldots,s_n)$, and $\mathbb{S}=\left\{{\bf s}|{\bf s}\in\{0,1\}^n\right\}$ or the set of all possible states. For $j\in V$, we denote its parent node set by ${\bf pa}_j$, which is the set of nodes that point to $j$, namely ${\bf pa}_j=\{i:(i,j)\in E\}$. For example, the parent node set of node $5$ in Figure \ref{fig:graph-theoretic-representation} is ${\bf pa}_5=\{3, 7\}$.
{Let $S_{{\bf pa}_{j}}$ denote the states of the parent nodes of node $j$.}

\smallskip

\noindent{\bf Remark}.
We can use the preceding representation to enable ``what if'' analysis by accommodating zero-day vulnerabilities. For a hypothetical zero-day vulnerability, one needs to determine the probability it can be an entry point, and the conditional probability $e_{ij}$, where both $i$ and $j$ can be zero-day vulnerabilities. 

\subsection{Modeling Cyber Risks in Terms of Business Lines}

\noindent{\bf Modeling security states in a smart home}.
Table \ref{Tab:V-Loss-sh} summarizes the aforementioned impacts of the exploitation of the 7 vulnerabilities with respect to the 6 business lines, while recalling that the exploitation of one vulnerability can impact multiple business lines. 

\begin{table}[htbp!]
\begin{center}
\begin{tabular}{r|c|c|c|c|c|c|c|c} 
 \hline
Smart Home Device & Node Identity &Vulnerability & $\L_1 $ & $\L_2$& $\L_3 $ & $\L_4$& $\L_5 $ & $\L_6$  \\  \hline
iPhone & 1 & CVE-2022-22667  & $\checkmark$  & &   &  &$\checkmark$  &   \\ \hline
Smart TV & 2& CVE-2020-27403  & $\checkmark$  & &   &  &  &   \\ \hline
Smart Home Hub & 3& CVE-2018-3919   & $\checkmark$  &$\checkmark$ &    &  &  &    \\  \hline
{Smart Sensor} & 4 & CVE-2021-29438  & $\checkmark$  & &   &  &  &    \\ \hline
Smart Camera & 5 & CVE-2021-32934 &   & $\checkmark$&    & $\checkmark$ &   &   \\  \hline
Smart Lock & 6 & CVE-2019-7256  &   & &   &  &  & $\checkmark$  \\ \hline
Laptop & 7 & CVE-2017-8759 & $\checkmark$  & & $\checkmark$  &  &  &   \\ \hline
\end{tabular}
\caption{Vulnerabilities in a smart home and losses caused by their exploitation, where $\checkmark$ indicates applicability,
vulnerability $i\in \{1,\ldots,7\}$, and business line $ j\in \{1, \ldots, 6\}$.\label{Tab:V-Loss-sh}}
\end{center}
\end{table} 

Let $\mathcal{L}_m$ denote the subset of nodes in $V$ whose exploitation can incur loss in business line $m$, namely $\L_m$, where $m\in \{1,\ldots,M\}$. According to Table \ref{Tab:V-Loss-sh}, we have $\mathcal{L}_1=\{1,2,3,4,7\}$, $\mathcal{L}_2=\{3,5\}$, $\mathcal{L}_3=\{7\}$, $\mathcal{L}_4=\{5\}$, $\mathcal{L}_5=\{1\}$, and $\mathcal{L}_6=\{6\}$.
Then, we have \cite{koller2009probabilistic}:
\vspace{-1em}
\begin{equation}\label{eq:joint}
\Pr({\bf S}={\bf s})=\Pr(S_{1}=s_{1},\ldots, S_{n}=s_{n})=\prod_{i=1}^{n}\Pr(S_{i}=s_{i}|{S_{{\bf pa}_{i}}}),~ s_{i}\in\{0,1\}.
\end{equation}
\ignore{
where 
$$s_{i}=
\left\{
\begin{array}{rl}
  1,&  \mbox{the corresponding device is compromised,}   \\
 0, &  \mbox {the corresponding device is not compromised.}
  \end{array}
\right.
 $$
}
For the loss incurred in business line $m\in \{1,\ldots,M\}$, we have
\begin{eqnarray}\label{eq:loss-dep1}
\Pr\left(\L_m \leq \ell_m\right) 
&=&
{\sum_{{\bf s}\in \mathbb{S}}} \Pr\left( {\L}_m\le \ell_m|{\bf S}={\bf s} \right)
\Pr({\bf S}={\bf s})\nonumber\\
&=&{\sum_{{\bf s}\in \mathbb{S}}}  \Pr\left( {\L}_m\le \ell_m|{\bf S}={\bf s} \right)  \prod_{i=1}^{n}\Pr(S_{i}=s_{i}|{S_{{\bf pa}_{i}}}).
\end{eqnarray}
Then, we have
\vspace{-1em}
\begin{eqnarray}\label{eq:loss-total1}
\Pr\left({\TL}\le \ell\right)&=&
\Pr\left(\sum_{m=1}^M {\L}_m\le \ell \right)\nonumber\\\
&=&
{\sum_{{\bf s}\in \mathbb{S}}} \Pr\left(\sum_{m=1}^M {\L}_m\le \ell|{\bf S}={\bf s}\right) \prod_{i=1}^{n}\Pr(S_{i}=s_{i}|{S_{{\bf pa}_{i}}}).
\end{eqnarray}
{Note that we cannot compute $\Pr(\L_m\leq \ell_m)$ and $\Pr(\TL\leq \ell)$ according to Eqs.\eqref{eq:loss-dep1} and  \eqref{eq:loss-total1}  
without knowing both $S_{{\bf pa}_i}$ and the {dependence} among the $\L_m$'s.
The issue of $S_{{\bf pa}_i}$ could be resolved if we can assume that the exploitation of $j$ is independently incurred by its compromised parent nodes $i \in {\bf pa}_{j}$, which means that the exploitation probability of $j\in V$ can be computed as 
\begin{equation}\label{eq:exploit}
   \Pr(S_{j}=1|{S_{{\bf pa}_{j}}})=\left\{
    \begin{array}{cc}
         0 &\forall i \in {\bf pa}_{j}, S_i=0; \\
         1-\prod_{i \in {\bf pa}_{j}, S_i=1}(1- e_{ij}), & \mbox{otherwise} 
    \end{array}
    \right..
    \end{equation}
However, the computation would be very complex due to the BAG structure. 
Moreover, Eq.\eqref{eq:loss-total1} is still infeasible to compute because of the dependence among the $\L_m$'s.    
To tackle these two issues,
we propose using Algorithm \ref{alg:alg1} to conduct the simulation to empirically derive $S_{{\bf pa}_i}$ and the total loss without computing the {dependence} among the $\L_m$'s.} \\

\noindent{\bf Running example of modeling security states in a smart home}. Consider the example BAG in Figure \ref{fig:sh_new}. Eq.\eqref{eq:joint} says 
\begin{eqnarray}\label{Eq:joint-V}
\Pr(\mathbf{S}={\bf s})&=&\Pr(S_1=s_{1})\cdot \Pr(S_2=s_{2})\cdot \Pr(S_7=s_{7}) \nonumber\\
&&\cdot \Pr(S_3=s_{3}|S_2=s_{2},S_1=s_{1})\cdot \Pr(S_5=s_{5}|S_3=s_{3},S_7=s_{7}) \nonumber\\
&&\cdot \Pr(S_4=s_{4}|S_3=s_{3})\cdot \Pr(S_6=s_{6}|S_4=s_{4},S_7=s_{7}).
\end{eqnarray}
Suppose $S_{1}=1$ and $S_{2}=1$. Recall that $e_{23}=\Pr(S_3=1|S_2=1)$ and $e_{13}=\Pr(S_3=1|S_1=1)$. Then, we have
$$\Pr(S_3=1|S_2=1,S_1=1)=1-(1-e_{13})(1-e_{23}).$$ 
Since there are 7 vulnerabilities, there are $2^{7}$ possible states in $\mathbb{S}$.
Table \ref{Tab:V-P-sh} presents the probabilities of 8 (out of the $2^7$) states, computed according to Eq.\eqref{Eq:joint-V}. 

\begin{table}[htbp!]
\begin{center}
\begin{tabular}{ccccccc|c} 
 \hline
 $S_1$&  $S_2$&  $S_3$&  $S_4$& $S_5$&  $S_6$& $S_7$ & $Prob$ \\  \hline
 0 & 0 & 0 & 0 & 0 & 0 & 0 &  .097  \\ \hline
 0 & 0 & 0 & 0 & 0 & 0 & 1 &  .856  \\  \hline
 0 & 0 & 0 & 0 & 0 & 1 & 0 & .000 \\  \hline
 0 & 0 & 0 & 0 & 0 & 1 & 1 & .009  \\  \hline
 0 & 0 & 0 & 0 & 1 & 0 & 0 & .000  \\  \hline
 0 & 0 & 0 & 0 & 1 & 0 & 1 & .009 \\  \hline
 0 & 0 & 0 & 0 & 1 & 1 & 0 & .000 \\  \hline
 0 & 0 & 0 & 0 & 1 & 1 & 1 & .000 \\  \hline
\end{tabular}
\caption{Probability ($Prob$) of the smart home in 8 example states}
\label{Tab:V-P-sh}
\end{center}
\end{table}


\ignore{ 
\begin{table}[htbp!]
\begin{center}
\begin{tabular}{c|ccccccc|c} 
 \hline
 Scenario & $S_1$&  $S_2$&  $S_3$&  $S_4$& $S_5$&  $S_6$& $S_7$ & $Prob$ \\  \hline
1 & 0 & 0 & 0 & 0 & 0 & 0 & 0 &  .097  \\ \hline
2 & 0 & 0 & 0 & 0 & 0 & 0 & 1 &  .856  \\  \hline
3 & 0 & 0 & 0 & 0 & 0 & 1 & 0 & .000 \\  \hline
4 & 0 & 0 & 0 & 0 & 0 & 1 & 1 & .009  \\  \hline
5 & 0 & 0 & 0 & 0 & 1 & 0 & 0 & .000  \\  \hline
6 & 0 & 0 & 0 & 0 & 1 & 0 & 1 & .009 \\  \hline
7 & 0 & 0 & 0 & 0 & 1 & 1 & 0 & .000 \\  \hline
8 & 0 & 0 & 0 & 0 & 1 & 1 & 1 & .000 \\  \hline
\end{tabular}
\caption{$Prob$ represents the joint probabilities of vulnerabilities in different scenarios.}
\label{Tab:V-P-sh}
\end{center}
\end{table}
}

{Now we show how to apply Algorithm \ref{alg:alg1} to empirically compute the $\L_m$'s and thus the $\TL$ based on reasonable assumptions.}
Suppose $\L_1$ and $\L_{2}$ follow the exponential  distribution, $\L_1 \sim \exp(\lambda_1)$ and $\L_2 \sim \exp(\lambda_2)$, where  
$\lambda_1=\sum_{i\in \mathcal{L}_1}  {\bm a}_1 S_i$, $\lambda_2=\sum_{j\in \mathcal{L}_2} {\bm a}_2 S_j$, {${\bm a}_1=(1/160,1/32,1/80,1/80,1/160)$ is the coefficients representing the rates of loss incurred by the exploitation of vulnerabilities in $\mathcal{L}_1=\{1,2,3,4,7\}$ with the rate of loss
incurred by the exploitation of vulnerability $1$ being $1/160$
per year (i.e., the mean loss is \$160 per year), which is an input parameter that can be estimated based on historic data or domain expertise, 
and ${\bm a}_2=(1/640, 1/320)$ 
is the coefficients representing the rates of loss incurred by the exploitation of vulnerabilities in $\mathcal{L}_2=\{3,5\}$.} The exponential distribution assumptions about $\L_1$ and $\L_{2}$ can be justified as follows: { For a smart home, $\L_1$ (data breach) and $\L_2$ (loss of use) may not be extremely large losses, and thus can be modeled as an exponential distribution \cite{casella2024statistical}.}

Suppose $\L_{3}$ and $\L_{4}$ follow lognormal distributions when the vulnerabilities in $\mathcal{L}_3=\{7\}$ and $\mathcal{L}_4=\{5\}$ are exploited, namely ${\L}_{3}\sim {\rm Lognorm}\left(\mu_1,\sigma^{2}\right)$ and ${\L}_{4}\sim {\rm Lognorm}\left(\mu_2,\sigma^{2}\right)$, where $(\mu_1, \mu_2, \sigma)=(4, 7, 1)$ {based on historic data or domain expert's estimation}.  The lognormal distribution assumption can be justified as follows: For a smart home, $\L_3$ (ransomware) and $\L_4$ (cyber extortion) {could be large losses and thus can be modeled via the heavy tail of the lognormal distribution \cite{casella2024statistical}.} 


Suppose $\L_{5}$ and $\L_{6}$ follow Gamma distributions when the vulnerabilities in $\mathcal{L}_5=\{1\}$ and $\mathcal{L}_6=\{6\}$ are exploited, namely ${\L}_{5}\sim \Gamma\left(\alpha_1,\beta\right)$ and ${\L}_{6}\sim \Gamma\left(\alpha_2,\beta\right)$, where $(\alpha_1,\alpha_2,\beta)=(1000,2000,1)$ {based on historic data or domain expert's estimation}. The Gamma distribution assumption can be justified as follows: {For a smart home, the loss incurred by $\L_5$ (online fraud) and $\L_6$ (property theft) could vary significantly and thus can be modeled by the flexibility of the Gamma distribution as it has two parameters (i.e., the shape parameter and the scale parameter) \cite{casella2024statistical}.}

Table \ref{Tab:Sum-sh-pricing} summarizes the simulation result of $\L_m$ for $m\in \{1,\ldots,6\}$ and the resulting total loss $\TL$, where $n=7$ and $R=10,000$.
We observe that ${\L_1}$ exhibits the highest mean value because vulnerability $7$ is most likely exploited,
that ${\L_3}$ has the second highest mean value,
and ${\L_4}$ exhibits extreme values because cyber extortion could cause significant loss when the smart camera is exploited. 

\begin{table}[htbp!]
\small
\centering
 \resizebox{\textwidth}{!}{ \begin{tabular}{c|c|c|c|c|c|c|c|c|c|c|c|c}
 \hline 
    & Min & $Q_{25}$ & Median & $Q_{75}$ & $Q_{90}$ & $Q_{95}$ & $Q_{99}$ & $Q_{99.5}$ & $Q_{99.9}$& Max & Mean & SD\\
    \hline
    $\L_1$ & .00 & 28.31 & 92.35 & 204.10 & 354.61 & 476.86 & 749.93 & 867.01 & 1143.49 & 1875.83 & 144.81 & 165.50 \\ 
    $\L_2$ & .00 & .00 & .00 & .00 & .00 & .00 & .00 & 199.73 & 704.06 & 1388.21 & 3.02 & 43.15 \\ 
    $\L_3$ & .00 & 21.26 & 48.13 & 99.59 & 191.60 & 273.29 & 57.05 & 727.44 & 1047.45 & 4597.62 & 83.16 & 123.43 \\ 
    $\L_4$ & .00 & .00 & .00 & .00 & .00 & .00 & .00 & 868.25 & 4867.59 & 15563.73 & 18.80 & 319.33 \\ 
    $\L_5$ & .00 & .00 & .00 & .00 & .00 & .00 & .00 & 998.13 & 1029.31 & 1069.61 & 9.46 & 96.69 \\ 
    $\L_6$ & .00 & .00 & .00 & .00 & .00 & .00 & .00 & 2004.76 & 2072.27 & 2150.52 & 19.70 & 198.09 \\ \hline
  $\TL$ & .00 & 87.22 & 181.37 & 332.07 & 539.64 & 771.11 & 2142.00 & 2458.68 & 5326.47 & 16521.96 & 278.95 & 465.62 \\ 
    \hline 
\end{tabular}}
\caption{Simulation result based on Algorithm \ref{alg:alg1} and parameters discussed in the text\label{Tab:Sum-sh-pricing}}
\end{table}




\subsection{Determining Cyber Insurance Premiums and Deductibles}
\label{sec:pricing}
\label{sec:task3}

\noindent{\bf Is the current real-world smart home insurance policy competent?}
We use the term {\em competency} to indicate that a smart home cyber insurer can make a profit and a smart home owner does not overly pay premium or deductible. 
We consider two real-world insurers whose names are anonymized in this paper: Insurer A requiring deductibles vs. Insurer B not requiring deductibles.

\underline{\bf Case 1: Insurer A requiring deductibles}. In this case, the one-year policy has a \$1,000 deductible and a \$50,000 coverage limit. The coverage includes cyber extortion ($\L_4$), data restoration, crisis management, and cyber bullying, {where the last three are not considered in this paper}. Table \ref{Tab:AIG} summarizes the yearly premium per {business line}.

\vspace{-1em}

\begin{table}[htbp!]
\scriptsize
    \centering
    \begin{tabular}{c|c|c|c|c|c}
    \hline
         Type &  {\bf Cyber extortion} & Data restoration & Crisis management  & Cyber bullying & Total Premium \\
         \hline
         Premium & 28 & 151 & 231 & 28 & 438\\
         \hline
    \end{tabular}
    \caption{Cyber insurance premium (\$) offered by Insurer A.
    \label{Tab:AIG}}
\end{table}

\vspace{-2em}

Since $\L_4$ is common to Insurer A's policy and the present study, we use it as a { baseline} to determine the parameters in our pricing formulas. Specifically, we set the premium for $\L_4$ to {\$28 per year}, {and the deductible and coverage limit to the same as that of Insurer A's.} Then, we determine the parameters in $\rho_1$ to $\rho_4$ based on the formulas given in the framework (Section \ref{sec:premium}) and the simulated losses. The { resulting parameters are $\theta=.5, .03, .25$ for $\rho_1, \rho_2,\rho_3$, and $\beta=.34$ for $\rho_4$,} 
respectively.

\vspace{-1em}  
\begin{table}[htbp!]
    \centering
    \begin{tabular}{c|c|c|c|c}
    \hline 
  Premium (\$)  & $\rho_1$ & $\rho_2$ & $\rho_3$ & $\rho_4$ \\
    \hline
  $\L_1$ & 217 & 150 & 185 & 211 \\ 
  $\L_2$ & 5 & 4 & 5 & 5  \\ 
  $\L_3$ & 125 & 87 & 107 & 120 \\ 
  $\L_4$ & {\bf 28} & {\bf 28}  & {\bf 28}  & {\bf 28}   \\ 
  $\L_5$ & 14 & 12 & 14 & 14 \\ 
  $\L_6$ & 30 & 26 & 29 & 30  \\ 
  \hline
Total premium  & 418 & 307 & 368 & 408  \\ 
  \hline 
    \end{tabular}
\caption{Our premiums (\$) under the 4 pricing principles, where `total premium' is the sum of all premiums.
 \label{table:pre}}
\end{table}
\vspace{-2em}

Table \ref{table:pre} shows the premium for each business line under the 4 premium principles described in the framework, where the total premium is the sum of individual premiums in the 6 business lines or $\sum_{m=1}^6 \rho_j(\L_m)$ where $j\in\{1,\ldots, 4\}$. 
We observe $\rho_{1}$ is the largest total premium (\$418) and $\rho_{2}$ is the smallest (\$307). 

To assess the performance of premium pricing principles, we use the aforementioned two metrics: 
Profit, namely $\mbox{Profit}=\mbox{Premium} - \mbox{Claim}$; and Loss Ratio (LR), namely
$\mbox{LR}= \mbox{Claim}/\mbox{Premium}$.
Suppose the permissible LR is 40\%, and
500 smart home owners purchase smart home insurance (i.e., a portfolio of 500 policyholders) where premiums are charged according to Table \ref{table:pre}.  {We simulate the loss scenarios of the portfolio with 10,000 independent runs and consider the distribution of the loss scenarios of these 10,000 simulation runs}.  

\vspace{-1em}
 \begin{table}[htbp!]
    \centering
  \resizebox{.98\textwidth}{!}{  
\begin{tabular}{ c|l|cccccccc|cc}
    \hline 
        & & Min &  $Q_1$ & $Q_5$ &  $Q_{10}$ &  $Q_{15}$ &  $Q_{50}$ & $Q_{75}$ &  Max & Mean & SD \\
  \hline
$\rho_1$  & Profit & 127,549 & 174,542 & 183,192 & 186,869 & 189,152 & 196,215 & 199,612 & 207,628 & 195,089 & 6,429 \\ 
    \hline
$\rho_2$  & Profit  & 72,049 & 119,042 & 127,692 & 131,369 & 133,652 & 140,715 & 144,112 & 152,128 & 139,589 & 6,429 \\ 
    \hline
$\rho_3$  & Profit  & 102,549 & 149,542 & 158,192 & 161,869 & 164,152 & 171,215 & 174,612 & 182,628 & 170,089 & 6,429 \\ 
    \hline
$\rho_4$ &Profit  & 122,549 & 169,542 & 178,192 & 181,869 & 184,152 & 191,215 & 194,612 & 202,628 & 190,089 & 6,429 \\ 
         \hline \hline 

        & & Min &  $Q_{25}$ & $Q_{50}$ &  $Q_{75}$  &  $Q_{90}$ & $Q_{95}$ & $Q_{99.5}$ &   Max & Mean & SD \\
  \hline
$\rho_1$  & LR & .01 & .04 & .06 & .08  & .11 & .12 & .19  & .39 & .07 & .03 \\ 
    \hline
$\rho_2$  & LR & .01 & .06 & .08 & .11  & .14 & .17 & .26  & .53 & .09 & .04 \\ 
    \hline
$\rho_3$  & LR& .01 & .05 & .07 & .09  & .12 & .14 & .22  & .44 & .08 & .03 \\ 
    \hline
 $\rho_4$  & LR & .01 &.05 & .06 & .08  & .11 & .13 & .19  & .40 & .07 & .03 \\ 
         \hline 
    \end{tabular}
    }
\caption{Summary statistics of Profits and LRs under the 4 pricing principles with \$1,000 deductible and \$50,000 coverage limit, where `Profit' and `LR' correspond to the loss of individual business lines (i.e., total premium in Table \ref{table:pre}). 
\label{Tab:AIG-base}}
 \end{table}
\vspace{-2em}

Table \ref{Tab:AIG-base} shows the summary statistics of portfolio Profit and LRs based on the loss of individual business lines and the aggregated loss under the 4 pricing principles. We observe that all the profits are positive, with $\rho_{1}$ being the largest and $\rho_{2}$ being the smallest. Moreover, the mean LRs are small ($<.1$) for every pricing principle, the high quantiles of LRs (e.g., $Q_{99.5}$) are still smaller than $40\%$, but the worst-case scenarios for $\rho_2$ and $\rho_3$ are beyond the permissible LR of 40\%. 
Note that all the standard deviations are the same because the coverage limits and deductibles are fixed.

\begin{insight}
Insurer A, which requires deductibles, is too conservative, meaning that home owners 
are over charged for their smart home cyber insurance.
\end{insight}

\underline{\bf Case 2: Insurer B not requiring deductibles}. In this case,
Insurer B offers a smart home policy covering cyber extortion and ransomware, cyber financial loss, and cyber personal protection with coverage limits and premiums shown in Table \ref{Tab:Chubb}.   

\vspace{-1em}
\begin{table}[htbp!]
 \scriptsize
    \centering
    \begin{tabular}{c|c|c|c|c}
    \hline 
    \multicolumn{4}{c|}{Coverage limit}  & \multirow{2}{*}{Premium}\\
    \cline{1-4}
      Cyber extortion & Cyber financial loss & Cyber personal protection & All covered events  & \\
       \hline
        10,000 & 50,000 & 50,000 & 50,000 & 200 \\
         \hline 
    \end{tabular}
\caption{Smart home insurance policy offered by Insurer B with 0 deductible and \$50,000 coverage limit for the covered attacks.}
    \label{Tab:Chubb}
\end{table}

\vspace{-2em}


We apply the premium strategy of Insurer B to our simulated portfolio losses, and present the summary statistics of the Profit and LR in  Table \ref{Tab:Chubb-P-LR}. We observe that under this premium strategy, Insurer B cannot make a profit, and 
the mean LR is 1.35 which is much larger than the permissible LR (40\%).

\vspace{-1em}

\begin{table}[htbp!]
    \centering
  \resizebox{\textwidth}{!}{   \begin{tabular}{c|ccccccccc|cc}
    \hline 
        & Min &  $Q_{1}$ & $Q_{5}$ &  $Q_{10}$ &  $Q_{50}$ &  $Q_{95}$ & $Q_{995}$ & $Q_{999}$ &  Max & Mean & SD \\
        \hline
       Profit &  -106,801 & -61,674 & -51,710 & -46,830 & -34,010 & -20,912 & -13,866 & -10,635 & -4,540 & -34,764 & 9,416 \\
       \hline
       LR & 1.05 & 1.16 & 1.21 & 1.24 & 1.34 & 1.52 & 1.65 & 1.74 & 2.07 & 1.35 & .09 \\ 
         \hline 
    \end{tabular}}
\caption{Summary statistics of profit and LR under premium practice of Insurer B.    \label{Tab:Chubb-P-LR}}
\end{table}

\vspace{-3em}

\begin{insight}
Insurer B, which does not require deductibles, cannot make a profit and cannot survive.
\end{insight}

\noindent{\bf Our proposals for competent smart home insurance policies}.
{Given that the current real-world smart home cyber insurance policies are not competent, we propose our insurance policy that would be more competent (i.e., benefit both insurers and smart home owners). Since requiring no deductible is not a standard practice and is not profitable in our analysis mentioned above, we only consider the case of requiring deductibles. We ask and address two questions: Given the same premium and coverage limit as Insurer A in practice, what is the competent (or affordable) deductible that makes the insurer profitable? Given the same deductible and coverage limit as Insurer A, what is the competent 
 (or affordable) premium that makes the insurer profitable?}

\underline{\bf Seeking small yet competent deductibles}. To search for competent insurance polices, or small deductibles, we start the \$1,000 required by Insurer A while making the premiums and coverage limit the same as that of Insurer A. We consider two strategies: (i) the permissible mean LR is 40\%;
(ii) the permissible high quantile of LR, $Q_{99.5}$, is 40\%. {We consider these two strategies because they reflect an insurer's risk attitude from the policy-making perspective. Strategy (i) ensures that, on average, the insurer maintains a profitable margin while providing coverage. Strategy (ii) represents a risk-averse approach, ensuring that the insurer remains profitable even in extreme scenarios.}

We start with a \$500 deductible, which would be more attractive to smart home owners than Insurer A which requires a $\$1,000$ deductible.

\vspace{-1.5em}
\begin{table}[htbp!]
    \centering
 \resizebox{\textwidth}{!}{       \begin{tabular}{l|l|cccccccc|cc}
    \hline\hline
        & & Min &  $Q_1$ & $Q_5$ &  $Q_{10}$ &  $Q_{15}$ &  $Q_{50}$ & $Q_{75}$ &  Max & Mean & SD \\
\hline
 $\rho_1$  &Profit & 112,217 & 157,355 & 166,918 & 170,996 & 173,536 & 181,815 & 186,350 & 199,217 & 180,880 & 7,743 \\ 
  \hline
 $\rho_2$   &Profit  & 56,717 & 101,855 & 111,418 & 115,496 & 118,036 & 126,315 & 130,850 & 143,717 & 125,380 & 7,743 \\ 
  \hline
 $\rho_3$ & Profit  & 87,217 & 132,355 & 141,918 & 145,996 & 148,536 & 156,815 & 161,350 & 174,217 & 155,880 & 7,743 \\ 
  \hline 
 $\rho_4$ & Profit  & 107,217 & 152,355 & 161,918 & 165,996 & 168,536 & 176,815 & 181,350 & 194,217 & 175,880 & 7,743 \\ 
  \hline\hline

        & & Min &  $Q_{25}$ & $Q_{50}$ &  $Q_{75}$  &  $Q_{90}$ & $Q_{95}$ & $Q_{99.5}$ &  Max & Mean & SD \\
\hline
 $\rho_1$   & LR  & .05 & .11 & .13 & .15 & .18 & .20 & .27  & .46 & .13 & .04 \\ 
  \hline
 $\rho_2$  &  LR  & .06 & .15 & .18 & .21  & .25 & .27 & .37  & .63 & .18 & .05 \\ 
  \hline
 $\rho_3$  &  LR & .05 & .12 & .15 & .18  & .21 & .23 & .31 & .53 & .15 & .04 \\ 
  \hline
 $\rho_4$   &  LR  & .05 & .11 & .13 & .16  & .19 & .21 & .28  & .47 & .14 & .04 \\ 
         \hline 
    \end{tabular}}
\caption{Summary statistics of Profits and LRs with \$500 deductible and \$50,000 coverage limit.
}
  \label{Tab:d500}
\end{table}
\vspace{-2.5em}



Table \ref{Tab:d500} presents the resulting Profits and LRs with a deductible of \$500. 
We observe that Profit of the insurer 
decreases when compared with that of Insurer A (Table \ref{Tab:AIG-base}), but this decrease is still acceptable because the mean LR is far below 40\% for \(\rho_1\) to \(\rho_4\), which is required by strategy (i). 
The same conclusion applies to strategy (ii), which requires that \(Q_{99.5}\) is 40\%.

The fact that a \$500 deductible is profitable in both strategies (i) and (ii) prompts us to {search for the strategy that is profitable to insurers while requiring a deductible that is as small as possible. Specifically, we gradually reduce the quantity of deductibles under each strategy}, as Eq. \eqref{eq:remain} indicates 
that the loss faced by the insurer decreases with the amount of deductible. Details follow.

First, consider a \$250 deductible. Owing to space limit, we omit the details but highlight the result as follows. We find that the Profit decreases.
However, under strategy (i), the mean LR is still less than 40\%, indicating that we can continue to decrease the deductible. Under strategy (ii), the \(Q_{99.5}\) of LR for \(\rho_1\) is equal to 40\%, meaning a \$250 deductible is sufficient; however, the \(Q_{99.5}\) of LR for \(\rho_2\), \(\rho_3\), and \(\rho_4\) exceeds 40\%, meaning a \$250 deductible is not sufficient.

Second, consider a \$200 deductible.
While omitting the details (owing to space limit), we highlight we observe a further decrease in Profit, which is expected.
Under strategy (i), we observe the mean LR for \(\rho_2\) surpasses 40\%, meaning a \$200 deductible is not sufficient.
Under strategy (ii), the \(Q_{99.5}\) of LR under all the 4 pricing principle exceeds 40\%, meaning a \$200 deductible is not enough.

Third, consider a \$150 deductible. 
While omitting details (owing to space limit), we highlight that the Profit further decrease.
Under strategy (i), the mean LR for \(\rho_3\) surpasses 40\%, meaning a \$150 deductible is not sufficient; the mean LR for 
\(\rho_1\) and \(\rho_4\) 
is less than 40\%, meaning a smaller deductible is possible. Under strategy (ii), a \$150 deductible is also not profitable.

\vspace{-1.5em}
\begin{table}[htbp!]
    \centering
    \resizebox{\textwidth}{!}{         \begin{tabular}{l|l|cccccccc|cc}
    \hline 
        & & Min &  $Q_1$ & $Q_5$ &  $Q_{10}$ &  $Q_{15}$ &  $Q_{50}$ & $Q_{75}$ &  Max & Mean & SD \\
        \hline
    $\rho_1$ & Profit & 44,965 & 89,249 & 99,248 & 104,001 & 107,026 & 116,589 & 122,186 & 144,683 & 115,821 & 9,223 \\ 
          \hline
       $\rho_2$   & Profit & -10,535 & 33,749 & 43,748 & 48,501 & 51,526 & 61,089 & 66,686 & 89,183 & 60,321 & 9,223 \\ 
          \hline
         $\rho_3$  & Profit & 19,965 & 64,249 & 74,248 & 79,001 & 82,026 & 91,589 & 97,186 & 119,683 & 90,821 & 9,223 \\ 
          \hline
     $\rho_4$   & Profit & 39,965 & 84,249 & 94,248 & 99,001 & 102,026 & 111,589 & 117,186 & 139,683 & 110,821 & 9,223 \\ 
         \hline \hline 

        & & Min &  $Q_{25}$ & $Q_{50}$ &  $Q_{75}$  &  $Q_{90}$ & $Q_{95}$ & $Q_{99.5}$ &  Max & Mean & SD \\
        \hline
     $\rho_1$  & LR & .31 & .42 & .44 & .47  & .50 & .53 & .59  & .78 & .45 & .04 \\ 
          \hline
    $\rho_2$   & LR & .42 & .57 & .60 & .64  & .68 & .71 & .81  & 1.07 & .61 & .06 \\ 
          \hline
     $\rho_3$  & LR & .35 & .47 & .50 & .54 & .57 & .60 & .67  & .89 & .51 & .05 \\ 
          \hline
 $\rho_4$   & LR & .32 & .43 & .45 & .48  & .51 & .54 & .61  & .80 & .46 & .05 \\ 
         \hline 
    \end{tabular}}
\caption{Summary statistics of Profits and LRs with a \$100 deductible and \$50,000 coverage limit.
}
    \label{Tab:d100}
\end{table}

\vspace{-2.5em}

Fourth, consider a \$100 deductible. Table \ref{Tab:d100} presents the resulting Profits and LRs. We observe that the Profit further decreases. Under strategy (i), the mean LR under all the 4 pricing principles, including 
\(\rho_1\) and \(\rho_4\), which are profitable with a \$150 deductible, surpass 40\% under, meaning that a \$100 deductible is not sufficient. 
Under strategy (ii), a \$100 deductible is also not profitable.


Based on the preceding simulation results,
we propose the deductibles under the 4 pricing principles in Table \ref{Tab:d-result}, while showing their mean Profits. When compared with the real-world Insurer A's insurance policy
(Table \ref{Tab:AIG-base}), we observe that the mean Profits based on our insurance policies are reduced, {but insurers are still profitable, while smart home owners only need to pay a significantly smaller amount of deductibles, which would attract more smart owners}. More policyholders can increase the insurer's profit as they might face the same kinds of cyber risks.

\vspace{-1.5em}

\begin{table}[htbp!]
    \centering
   \resizebox{\textwidth}{!}{        \begin{tabular}{l|l|c|c||c|c||c|c}
    \hline 
        & \multicolumn{2}{c|}{Total premium}  & Coverage limit & Deductible 1& Mean Profit 1& Deductible 2& Mean Profit 2\\
        \hline
 $\rho_1$ & Premium & 418  & 50,000& 150&131,809 &250&154,670 \\
         \hline
        $\rho_2$ & Premium & 307 & 50,000& 250&99,170 &500&125,380\\
         \hline
        $\rho_3$ & Premium & 368 & 50,000& 200&119,583 &500&155,880 \\
         \hline
          $\rho_4$  &Premium & 408 & 50,000&150&126,809 &500 &175,880 \\
         \hline 
    \end{tabular}}
\caption{Our proposed deductibles, where
{Deductible 1 and Mean Profit 1 are derived based on strategy (i) or the permissible mean LR being smaller than 40\%, 
Deductible 2 and Mean Profit 2 are derived based on strategy (ii) or the permissible $99.5$th LR ($Q_{99.5}$) being smaller than 40\%}. 
}
    \label{Tab:d-result}
\end{table}

\vspace{-2.5em}

\begin{insight}
\label{insight:our-result-1}
When compared with the current smart home cyber insurance practice (via Insurer A's policy), our framework leads to smaller mean profits to cyber insurers but significantly smaller deductibles to home owners.
\end{insight}

\underline{\bf Seeking small yet competent premiums}. We fix the deductible at \$1,000 and the coverage limit at \$50,000 as in the policy of Insurer A.
We use the mean LR of 40\% in strategy (i) and the $Q_{99.5}$ LR of 40\% in strategy (ii),  to determine the respective premiums. 


\vspace{-1.5em}

\begin{table}[htbp!]
    \centering\small
 \resizebox{\textwidth}{!}{      \begin{tabular}{c|c|ccccccccc|cc}
    \hline 
   Premium & &  Min &  $Q_{1}$ & $Q_{5}$ &  $Q_{10}$ &  $Q_{50}$ &  $Q_{95}$ & $Q_{99.5}$ & $Q_{99.9}$ &  Max & Mean & SD \\
         \hline
     \multirow{2}{*}{{\bf 198}} &Profit & 17,549 & 64,542 & 73,192 & 76,869 & 86,215 & 93,089 & 95,761 & 96,699 & 97,628 & 85,089 & 6,429 \\
        &LR & .01 & .04 & .06 & .07 & .13 & .26 & {\bf .40} & .50 & .82 & .14 & .06 \\ 
        \hline
    \multirow{2}{*}{{\bf 70}} &Profit& -46,451 & 542 & 9,192 & 12,869 & 22,215 & 29,089 & 31,761 & 32,699 & 33,628 & 21,089 & 6,429 \\ 
    &LR  & .04 & .11 & .17 & .20 & .37 & .74 & 1.13 & 1.41 & 2.33 &{\bf  .40} & .18 \\ 
     \hline
        \hline 
    \end{tabular}}
\caption{Summary statistics of Profits and LRs with \$1,000 deductible and \$50,000 coverage limit.}
    \label{Tab:Pro-LR-Premium}
\end{table} 

\vspace{-2.5em}

Table \ref{Tab:Pro-LR-Premium} presents the premiums corresponding to the summary statistics of Profits and LRs. 
We observe that if the permissible LR of $Q_{99.5}$  is 40\%, {a \$198 premium leads to a \$85,089 mean profit (i.e., the mean of the total profit of the insurer in serving 500 smart homes); if the permissible mean LR is 40\%, a \$70 premium leads to a  \$21,089 mean profit.} Therefore, it is possible to charge a low premium (\$70) for a decent coverage (\$50,000), while remaining profitable. 

\begin{insight}
\label{insight:our-result-2}
When compared with the current smart home cyber insurance practice (via Insurer A's policy), our framework leads to a lower premium while insurers remain profitable.
\end{insight}


\section{Conclusion}
\label{sec:conclusion}
We have presented a framework for smart home cyber insurance that can lead to competent policies to make insurers profitable while smart home owners pay a more affordable premium or deductible than their counterpart in the current practice. 
We conducted case studies to demonstrate the usefulness of the framework, while showing that the current smart insurance pricing can be further adjusted to offer more attractive policies (e.g., lower deductible or premium). 

The present study has several limitations that need to be addressed in the future. First, the loss distributions and parameters used in our case study are assumed to be given. While they can be derived from
experience and/or historic data in principle,
this needs to be calibrated when real-world smart home cyber claim data are available. Second, our case study is based on BAG for analyzing the probability that a vulnerability in a smart home will be exploited. This method, or any method for the same purpose, needs to be validated with real-world smart home experiments. 
Third, we empirically searched for competent smart home cyber insurance deductibles and premiums. It is interesting to define optimal deductibles and premiums and solve such optimization problems analytically. Fourth, we do not consider the systemic risk that occurs when common vulnerabilities exist in multiple smart home networks. Fifth, we need to deepen our understanding of cyber risks in the business lines. For instance, characterizing and forecasting data breaches have been investigated at the enterprise or industry level \cite{XuTIFSSparsity2021,DBLP:journals/ejisec/FangXXZ19,XuJAS2018,XuTIFSDataBreach2018} but not at the smart home level ($\L_2$); characterizing the psychological aspects of cyber social engineering attacks has been conducted in a general context \cite{XuSciSec2024PF-Evolution,XuSciSec2024PTac-PTech-Evolution,XuPIEEE2024,XuFrontierInPsychology2020} but not the smart home context ($\L_5$).  
Sixth, it is interesting to extend or adapt the present study to accommodate other settings, such as the financial service and  healthcare sectors.

\smallskip

\noindent{\bf Acknowledgment}. We thank the reviewers for their comments. This research was supported in part by NSF Grant \#2115134 and Colorado State Bill 18-086. This research work is also a contribution to the International Alliance for Strengthening Cybersecurity and Privacy in Healthcare (CybAlliance, Project no. 337316).

\bibliographystyle{splncs04}

\begin{thebibliography}{10}
\providecommand{\url}[1]{\texttt{#1}}
\providecommand{\urlprefix}{URL }
\providecommand{\doi}[1]{https://doi.org/#1}

\bibitem{CVSS}
Common vulnerability scoring system.
  \url{http://www.rst.org/cvss/cvss-guide.html}, accessed: 2021-12-30

\bibitem{Nessus}
Nessus home page. \url{https://www.tenable.com/products/nessus}

\bibitem{OpenVAS}
Openvas home page. \url{https://www.openvas.org/}

\bibitem{bitdefender}
The 2024 iot security landscape report.
  \url{https://blogapp.bitdefender.com/hotforsecurity/content/files/2024/06/2024-IoT-Security-Landscape-Report_consumer.pdf}

\bibitem{Zion}
Zion market research report.
  \url{https://www.zionmarketresearch.com/report/smart-home-market} (2023)

\bibitem{8835392}
Alrawi, O., Lever, C., Antonakakis, M., Monrose, F.: Sok: Security evaluation
  of home-based iot deployments. In: 2019 IEEE Symposium on Security and
  Privacy (SP). pp. 1362--1380 (2019). \doi{10.1109/SP.2019.00013}

\bibitem{awiszus2023modeling}
Awiszus, K., Knispel, T., Penner, I., Svindland, G., Vo{\ss}, A., Weber, S.:
  Modeling and pricing cyber insurance: Idiosyncratic, systematic, and systemic
  risks. European Actuarial Journal  \textbf{13}(1),  1--53 (2023)

\bibitem{bohme2010modeling}
B{\"o}hme, R., Schwartz, G.: Modeling cyber-insurance: towards a unifying
  framework. In Proceedings of the Workshop on the Economics of Information
  Security  (2010)

\bibitem{valuepenguin}
Breiner, B.: What is personal cyber insurance? and how can homeowners buy a
  policy? \url{https://www.valuepenguin.com/personal-cyber-home-insurance}
  (2022)

\bibitem{bugeja2021prash}
Bugeja, J., Jacobsson, A., Davidsson, P.: Prash: A framework for privacy risk
  analysis of smart homes. Sensors  \textbf{21}(19), ~6399 (2021)

\bibitem{casella2024statistical}
Casella, G., Berger, R.: Statistical inference. CRC Press (2024)

\bibitem{XuSTRAM2018ACMCSUR}
Cho, J.H., Xu, S., Hurley, P.M., Mackay, M., Benjamin, T., Beaumont, M.: Stram:
  Measuring the trustworthiness of computer-based systems. ACM Comput. Surv.
  \textbf{51}(6),  128:1--128:47 (2019)

\bibitem{cremer2022cyber}
Cremer, F., Sheehan, B., Fortmann, M., Kia, A.N., Mullins, M., Murphy, F.,
  Materne, S.: Cyber risk and cybersecurity: a systematic review of data
  availability. The Geneva Papers on risk and insurance-Issues and practice
  \textbf{47}(3),  698--736 (2022)

\bibitem{das2011home}
Das, S.R., Chita, S., Peterson, N., Shirazi, B.A., Bhadkamkar, M.: Home
  automation and security for mobile devices. In: 2011 IEEE International
  Conference on Pervasive Computing and Communications Workshops (PERCOM
  Workshops). pp. 141--146. IEEE (2011)

\bibitem{denning2013computer}
Denning, T., Kohno, T., Levy, H.M.: Computer security and the modern home.
  Communications of the ACM  \textbf{56}(1),  94--103 (2013)

\bibitem{eling2022unraveling}
Eling, M., Jung, K., Shim, J.: Unraveling heterogeneity in cyber risks using
  quantile regressions. Insurance: Mathematics and Economics  \textbf{104},
  222--242 (2022)

\bibitem{DBLP:journals/ejisec/FangXXZ19}
Fang, X., Xu, M., Xu, S., Zhao, P.: A deep learning framework for predicting
  cyber attacks rates. {EURASIP} J. Information Security  \textbf{2019}, ~5
  (2019)

\bibitem{XuTIFSSparsity2021}
Fang, Z., Xu, M., Xu, S., Hu, T.: A framework for predicting data breach risk:
  Leveraging dependence to cope with sparsity. IEEE T-IFS  \textbf{16},
  2186--2201 (2021)

\bibitem{furman2019computing}
Furman, E., Kye, Y., Su, J.: Computing the gini index: A note. Economics
  Letters  \textbf{185},  108753 (2019)

\bibitem{furman2017gini}
Furman, E., Wang, R., Zitikis, R.: Gini-type measures of risk and variability:
  Gini shortfall, capital allocations, and heavy-tailed risks. Journal of
  Banking \& Finance  \textbf{83},  70--84 (2017)

\bibitem{hardy2006introduction}
Hardy, M.R.: An introduction to risk measures for actuarial applications. SOA
  Syllabus Study Note  \textbf{19} (2006)

\bibitem{he2024modeling}
He, R., Jin, Z., Li, J.S.H.: Modeling and management of cyber risk: a
  cross-disciplinary review. Annals of Actuarial Science pp. 1--40 (2024)

\bibitem{jacobs2021exploit}
Jacobs, J., Romanosky, S., Edwards, B., Adjerid, I., Roytman, M.: Exploit
  prediction scoring system (epss). Digital Threats: Research and Practice
  \textbf{2}(3),  1--17 (2021)

\bibitem{jacobsson2016risk}
Jacobsson, A., Boldt, M., Carlsson, B.: A risk analysis of a smart home
  automation system. Future Generation Computer Systems  \textbf{56},  719--733
  (2016)

\bibitem{kaas2008modern}
Kaas, R., Goovaerts, M., Dhaene, J., Denuit, M.: Modern actuarial risk theory:
  using R, vol.~128. Springer Science \& Business Media (2008)

\bibitem{klugman2012loss}
Klugman, S.A., Panjer, H.H., Willmot, G.E.: Loss Models: From Data to
  Decisions. Wiley (2012)

\bibitem{koller2009probabilistic}
Koller, D., Friedman, N.: Probabilistic graphical models: principles and
  techniques. MIT press (2009)

\bibitem{kozlov2012security}
Kozlov, D., Veijalainen, J., Ali, Y.: Security and privacy threats in iot
  architectures. In: Proceedings of the 7th International Conference on Body
  Area Networks. pp. 256--262 (2012)

\bibitem{lee2014securing}
Lee, C., Zappaterra, L., Choi, K., Choi, H.A.: Securing smart home:
  Technologies, security challenges, and security requirements. In: 2014 IEEE
  Conference on Communications and Network Security. pp. 67--72. IEEE (2014)

\bibitem{XuPIEEE2024}
Longtchi, T., Rodriguez, R.M., Al{-}Shawaf, L., Atyabi, A., Xu, S.:
  Internet-based social engineering psychology, attacks, and defenses: A
  survey. Proceedings of IEEE  \textbf{112}(3),  210--246 (2024)

\bibitem{XuSciSec2024PF-Evolution}
Longtchi, T., Xu, S.: Characterizing the evolution of psychological factors
  exploited by malicious emails. In: Proceedings of International Conference on
  Science of Cyber Security (SciSec'2024) (2024)

\bibitem{XuSciSec2024PTac-PTech-Evolution}
Longtchi, T., Xu, S.: Characterizing the evolution of psychological tactics and
  techniques exploited by malicious emails. In: Proceedings of International
  Conference on Science of Cyber Security (SciSec'2024) (2024)

\bibitem{ma2022frequency}
Ma, B., Chu, T., Jin, Z.: Frequency and severity estimation of cyber attacks
  using spatial clustering analysis. Insurance: Mathematics and Economics
  \textbf{106},  33--45 (2022)

\bibitem{marikyan2019systematic}
Marikyan, D., Papagiannidis, S., Alamanos, E.: A systematic review of the smart
  home literature: A user perspective. Technological Forecasting and Social
  Change  \textbf{138},  139--154 (2019)

\bibitem{XuFrontierInPsychology2020}
Montañez, R., Golob, E., Xu, S.: Human cognition through the lens of social
  engineering cyberattacks. Frontiers in Psychology  \textbf{11}, ~1755 (2020)

\bibitem{Pendleton16}
Pendleton, M., Garcia-Lebron, R., Cho, J.H., Xu, S.: A survey on systems
  security metrics. ACM Comput. Surv.  \textbf{49}(4),  62:1--62:35 (Dec 2016)

\bibitem{XuJAS2018}
Peng, C., Xu, M., Xu, S., Hu, T.: Modeling multivariate cybersecurity risks.
  Journal of Applied Statistics  \textbf{0}(0),  1--23 (2018)

\bibitem{poolsappasit2011dynamic}
Poolsappasit, N., Dewri, R., Ray, I.: Dynamic security risk management using
  bayesian attack graphs. IEEE Transactions on Dependable and Secure Computing
  \textbf{9}(1),  61--74 (2011)

\bibitem{sun2021modeling}
Sun, H., Xu, M., Zhao, P.: Modeling malicious hacking data breach risks. North
  American Actuarial Journal  \textbf{25}(4),  484--502 (2021)

\bibitem{sun2023multivariate}
Sun, H., Xu, M., Zhao, P.: A multivariate frequency-severity framework for
  healthcare data breaches. The Annals of Applied Statistics  \textbf{17}(1),
  240--268 (2023)

\bibitem{tasche2002expected}
Tasche, D.: Expected shortfall and beyond. Journal of Banking \& Finance
  \textbf{26}(7),  1519--1533 (2002)

\bibitem{XuUsenixSecurity2023}
Xia, Q., Chen, Q., Xu, S.: Near-ultrasound inaudible trojan (nuit): Exploiting
  your speaker to attack your microphone. In: Calandrino, J.A., Troncoso, C.
  (eds.) 32nd {USENIX} Security Symposium, {USENIX} Security 2023, Anaheim, CA,
  USA, August 9-11, 2023. {USENIX} Association (2023)

\bibitem{XuTIFSDataBreach2018}
Xu, M., Schweitzer, K.M., Bateman, R.M., Xu, S.: Modeling and predicting cyber
  hacking breaches. IEEE T-IFS  \textbf{13}(11),  2856--2871 (2018)

\bibitem{xu2019cybersecurity}
Xu, M., Hua, L.: Cybersecurity insurance: Modeling and pricing. North American
  Actuarial Journal  \textbf{23}(2),  220--249 (2019)

\bibitem{XuMTD2020}
Xu, S.: The cybersecurity dynamics way of thinking and landscape (invited
  paper). In: ACM Workshop on Moving Target Defense (2020)

\bibitem{XuCybersecurityDynamicsHotSoS14}
Xu, S.: Cybersecurity dynamics. In: Proceedings of the 2014 Symposium and
  Bootcamp on the Science of Security (HotSoS'14). p.~14. ACM (2014)

\bibitem{XuBookChapterCD2019}
Xu, S.: Cybersecurity dynamics: A foundation for the science of cybersecurity.
  In: Proactive and Dynamic Network Defense, vol.~74, pp. 1--31. Springer
  (2019)

\bibitem{XuSciSec2021SARR}
Xu, S.: Sarr: A cybersecurity metrics and quantification framework. In: Third
  International Conference on Science of Cyber Security (SciSec'2021). pp.
  3--17 (2021)

\end{thebibliography}

\end{document}